\documentstyle[12pt]{article}
\begin{document}
\def\a{\alpha}
\def\b{\beta}
\def\ch{\chi}
\def\d{\delta}
\def\e{\epsilon}
\def\E{{\cal E}}
\def\f{\phi}
\def\g{\gamma}
\def\h{\eta}  
\def\et{\tilde{\eta}}
\def\i{\iota}
\def\j{\psi}
\def\k{\kappa}
\def\l{\lambda}
\def\m{\mu}
\def\n{\nu}
\def\o{\omega}
\def\p{\pi}
\def\q{\theta}
\def\r{\rho}
\def\s{\sigma}
\def\t{\tau}
\def\u{\upsilon}
\def\x{\xi}
\def\z{\zeta}
\def\D{\Delta}
\def\F{\Phi}
\def\G{\Gamma}
\def\J{\Psi}
\def\L{\Lambda}
\def\O{\Omega}
\def\P{\Pi}
\def\S{\Sigma}
\def\U{\Upsilon}
\def\X{\Xi} 
\def\T{\Theta}
\def\vf{\varphi}
\def\ve{\varepsilon}
\def\cC{{\cal P}}
\def\cD{{\cal Q}}

\def\Ab{\bar{A}}
\def\gi{g^{-1}}
\def\li{{ 1 \over \l } }
\def\lb{\l^{*}}
\def\zb{\bar{z}}
\def\ub{u^{*}}
\def\vb{v^{*}}
\def\Tb{\bar{T}}
\def\pp {\partial }
\def\pb {\bar{\partial }}
\def\be{\begin{equation}}
\def\ee{\end{equation}}
\def\ben{\begin{eqnarray}}
\def\een{\end{eqnarray}}
\def\lt{\tilde{\lambda}}
\hsize=16truecm
\addtolength{\topmargin}{-1in}
\addtolength{\textheight}{1.5in}
\vsize=26truecm
\hoffset=-0.6in 
\baselineskip=7 mm

\thispagestyle{empty}
\begin{flushright} August \ 1997\\
SNUTP 97-082 \\
\end{flushright}
\begin{center}
 {\large\bf Matched Pulse Propagation in a Three-Level System  }
\vglue .5in
 Q-Han Park\footnote{ E-mail address; qpark@nms.kyunghee.ac.kr }
\vglue .2in
{and}
\vglue .2in
H. J. Shin\footnote{ E-mail address; hjshin@nms.kyunghee.ac.kr }
\vglue .2in
{\it  
Department of Physics \\
and \\
Research Institute of Basic Sciences \\
Kyunghee University\\
Seoul, 130-701, Korea}
\vglue .2in
{\bf ABSTRACT}\\[.2in]
\end{center}            
The B\"{a}cklund transformation for the three-level coupled 
Schr\"{o}dinger-Maxwell
equation is presented in the matrix potential formalism.
By applying the B\"{a}cklund transformation to a constant 
electric field background, we obtain a general solution for 
matched pulses (a pair of solitary waves) which can emit or 
absorb a light velocity solitary pulse but otherwise propagate 
with their shapes invariant. In the special case, this solution  
describes a steady state pulse without emission or absorption, 
and becomes the matched pulse solution recently obtained by Hioe 
and Grobe. A nonlinear superposition rule is derived from the 
B\"{a}cklund transformation and used for the explicit construction 
of two solitons as well as nonabelian breathers. Various new 
features of these solutions are addressed. In particular, we 
analyze in detail the scattering of ``invertons", a specific pair 
of different wavelength solitons one of which moving with the 
velocity of light. Unlike the usual case of soliton scattering, the 
broader inverton changes its sign through the scattering. 
Surprisingly, the light velocity inverton receives time 
advance through the scattering thereby moving faster than light, 
which however does not violate causality.

\vglue .1in
\newpage      
\section{Introduction}
The nonlinear interaction between radiation and a multi-level optical 
medium has received considerable interests for many years. Recently, 
this topic has attracted more attention in the context of lasing  
without inversion \cite{lwi,lwiexp} and electromagnetically induced 
transparency (EIT) \cite{Boller}. EIT is a technique for 
rendering an otherwise optically thick medium transparent to a 
weak probe laser by coupling the upper level coherently to a third 
level by a strong laser field. The transparency for pulses propagating 
through an optically thick medium has been known earlier, 
particularly for a soliton ($2 \pi $ pulse) in the shape of the 
hyperbolic-secant type  \cite{maim} and through a three-level 
medium \cite{bolshov}. More recently, there appeared exact analytic 
solutions for a pair of solitary waves, so called 
matched-solitary-wave pairs (MSP), propagating through a three-level 
medium whose invariant shapes are more general than the hyperbolic-secant 
type \cite{hioe}. 

In this paper, we present a new type of analysis for the three-level 
coupled Sch\"{o}dinger-Maxwell equation based on the matrix potential formalism. In 
particular, we find the B\"{a}cklund transformation of the coupled Sch\"{o}dinger-Maxwell 
equation in terms of the matrix potential variable and apply the 
B\"{a}cklund transformation to the constant electric field background
to obtain new MSP solutions which generalize the result 
in Ref. \cite{hioe}. These solutions in general describe the breakup of 
a MSP into another MSP with slower velocity and a soliton pulse moving 
with light velocity, or the reverse process of fusing a MSP and a soliton 
into another MSP. With a specific choice of parameters, these solutions 
reduce to the MSP solution in in Ref. \cite{hioe} which describes 
steady-state propogation of MSP without breakup. The generality in the 
shape of a MSP solution is explained through the $SU(2) \times U(1)$ group 
symmetry of the three-level $\L $-system with equal oscillator strengths.  
We show that the B\"{a}cklund transformation also allows a nonlinear 
superposition rule for solitons as well as MSP solutions. An explicit 
formula for the nonlinear superposition is given in terms of matrix 
potentials and used to generate two-soliton and nonabelian breather 
solutions. We consider in detail the scattering of a specific 
type of solitons which we call ``invertons". Invertons are a pair 
of different wavelength solitons one of which moving with the 
velocity of light and the other with slower velocity. Unlike the usual 
case of soliton scattering, the broader inverton changes its sign 
during the scattering process. Surprisingly, the light velocity inverton 
receives time advance through the scattering thereby moving faster 
than light. We show that however causality is not violated. 
A typical nonabelian breather describes a breathing $0 \pi$-pulse 
$E_{1} (v < c )$ which afterwards transfers to the nonbreathing 
$E_{2}$ pulse moving with light velocity.
                           
\section{Matrix potential formalism of the $\L$-system}

Consider a $\L $-configuration where level three is higher than level 
one and two. The system of equations governing the  propagation of 
pulses are given by the Schr\"{o}dinger equation
\ben
{\pp c_{1} \over \pp t} &=& i \O_{1} c_{3} 
\nonumber \\
{\pp c_{2} \over \pp t} &=& i \O_{2} c_{3} 
\nonumber \\ 
{\pp c_{3} \over \pp t} &=& i( \O_{1}^{*} c_{1} + \O_{2}^{*}c_{2} ) ,
\label{Sch} 
\een
and the Maxwell equation
\ben
i ({\pp \over \pp x} + {1 \over c}{\pp \over \pp t}) \O_{1} &=& 
l_{1} c_{1}c_{3}^{*} \nonumber \\
i ({\pp \over \pp x} + {1 \over c}{\pp \over \pp t}) \O_{2} &=& 
l_{2} c_{2}c_{3}^{*} .
\een
Here, $l_{i} =  2\pi N \m_{i}^{2} \o_{i} / \hbar ; ~ i =1,2 $ and 
$c_{k};~ k=1,2,3$ are slowly varying probability amplitudes for the 
level occupations, $\O_{i} = \m_{i} E_{i}/2 \hbar $ are the Rabi 
frequencies for the transitions $i \rightarrow 3$, 
$E_{1}$ and $E_{2}$ are the slowly varying electromagnetic field 
amplitudes,  $\m_{i}$ is the dipole matrix element for the relevant 
transition and $\o_{i}$ is the corresponding laser frequency, 
and $N$ is the density of resonant three-level atoms. For brevity, 
we introduce a new coordinate $ z = t - x/c, ~~ \zb = x/c $ so 
that  $ \pp \equiv \pp / \pp z = \pp / \pp t , ~~
\pb \equiv \pp / \pp \zb  = \pp / \pp t + c \pp / \pp x $.

As in our earlier papers \cite{park1,park2}, the main tool of our 
analysis will be using a matrix potential $g$ instead of the 
probability amplitudes $c_{i}$ in the following way; let $g$ be 
a $3 \times 3$ unitary matrix whose second row is the complex 
conjugation of probability amplitudes, i.e., 
\be
g = \pmatrix{ *& * & * \cr c_{1}^{*} & c_{2}^{*} & c_{3}^{*}  \cr 
 * & * & * }   ,  
\label{gmat}
\ee  
where the first and the third rows are to be determined later. In terms of 
$g$, the density matrix $\rho$ whose components are 
$\rho_{mk} = c_{m}c_{k}^{*}$ takes a simple form,
\be
\rho = {i \over l_{1}}  g^{-1} \Tb g, ~~~ 
\Tb = \pmatrix{0 &0 &0 \cr 0 & - i l_{1}  & 0 \cr 0 & 0 & 0} .
\label{density}
\ee
The specific choice of the matrix $\Tb$ is not essential. One may consider 
an arbitrary diagonal matrix $\Tb$ to handle density matrices in a more 
general context. Note that the first and the third rows of the matrix $g$ 
do not affect the density matrix $\rho $. In other words, the density matrix 
$\rho $ is invariant under the left multiplication of $g$ by 
any matrix $h$,
\be
g \rightarrow g^{'}=hg, 
\label{symm1}
\ee
which commutes with $\Tb$ and thus of the form
\be
h = \pmatrix{* & 0 & * \cr 0 & * & 0 \cr * & 0 & * } .  
\label{symm2}
\ee                               
At first sight, introducing the matrix $g$ with more redundant components 
than $c_{i}$'s  may seem an unnecessary complication. However, this is 
not so. In fact, it not only manifests the symmetry group structure of the
system, but it also simplifies the problem of solving differential 
equations.
Later, we show that the B\"{a}cklund transformation of the system, 
a solution generating technique, also takes a simple form in terms of $g$. 
The main advantage of using $g$ is that $g$ solves the Schr\"{o}dinger 
equation identically. In order to see this, we fix the redundancy 
introduced by Eqs. (\ref{symm1}) and (\ref{symm2}). Adopting the notation 
for the following matrix decomposition;
\be
\S = \S_{M} + \S_{H} , ~~ \S_{M}= \pmatrix{0 & 0 & * \cr 0 & 0 & * 
\cr * & * & 0}
, ~~ \S_{H} = \pmatrix{* & * & 0 \cr * & * & 0 \cr 0 & 0 & * },
\ee  
we fix the redundancy by imposing the constraint condition on $g$,
\be
(g^{-1}\pp g)_{H} = 0 , ~~ (\pb g g^{-1})_{H} = 0 .
\label{constr}
\ee
One can always solve the constraint by finding an $h$ which makes g to 
satisfy the constraint via the transform in 
Eq. (\ref{symm1}).\footnote{The existence of such an $h$ can be proved 
by adopting the field theory formulation of the problem as in 
\cite{park2}. However, we do not need the explicit expression of $h$. }
We also parameterize the remaining components of $g^{-1}\pp g$ such 
that\footnote{Note that $\gi \pp g $ is  anti-Hermitian and $g$ is unitary.}
\be
g^{-1} \pp g =  \pmatrix{ 0 & 0 & -i \O_{1} \cr      
0 & 0 & -i \O_{2} \cr - i\O_{1}^{*} & -i\O_{2}^{*} & 0}  .
\label{esch}
\ee                                                                    
The nonzero components in Eq. (\ref{esch}) express the Rabi frequencies 
in terms of $g$. In this new parametrization, one can readily check that 
the Schr\"{o}dinger equation (\ref{Sch}) arises from the simple identity,
\be
\pp  g^{\dagger} = \pp  g^{-1} = -g^{-1} \pp g g^{-1}  
= -g^{-1} \pp g g^{\dagger} .
\ee
This situation may be compared with the ordinary electromagnetism where 
the static electric field $\vec{E}$ in terms of a scalar potential $\phi $,
$\vec{E} = -\nabla \phi $, solves the curl-free condition 
$\nabla \times \vec{E} = 0$. Likewise, we solve the Schr\"{o}dinger equation 
in terms of a matrix potential $g$ and express the electric field components 
$\O_{i}$ in terms of $g$ as in Eq. (\ref{esch}).
Originally, the Schr\"{o}dinger and the Maxwell equations are made of 
five complex, component equations in total. Three of them 
(the Schr\"{o}dinger part) are now solved identically in terms of $g$, 
where $g$ is partially constrained by Eq. (\ref{constr}). Then, the remaining 
degree of $g$ can be parametrized by two unknown complex functions and the 
Maxwell equation changes into two component equations for these 
two unknown functions only.\footnote{ 
In mathematical terms, we have associated the density matrix $\rho $ with 
the coset $G/H= (SU(3)/SU(2) \times U(1))$ and introduced the matrix $g$ for 
the parametrization of $G/H$. The constraint equation on $g$ makes a specific 
choice for each equivalence class, and variables which parameterize 
equivalence classes are determined by the Maxwell equation.
Since the constraint restricts the subgroup $H (= SU(2) \times U(1))$-part of 
the variable $g$, the remaining unconstrained part of $g$ can be expressed  
in terms of two unknown complex functions $\varphi_{1}$ and $\varphi_{2}$. 
In the gauged sigma model formulation, the Maxwell equation is vector gauge 
invariant so that it decouples from the $H$-degree of freedom, i.e. it 
reduces to two complex equations only in $\varphi_{1}$ and $\varphi_{2}$. } 
In this way, the Maxwell equation resembles the Poisson equation in 
electrostatics. However, we do not need explicit component expressions in this 
paper so we supress them. What we need is the expression of the Maxwell 
equation in terms of $g$ such that
\be  
\pb (g^{-1} \pp g) = Q[T, ~  g^{-1} \Tb g]Q
\label{emaxe}
\ee
where
\be
T= \pmatrix{- {i\over 2} & 0 &0 \cr 
0 & -{i \over 2} & 0 \cr 0 & 0 & {i \over 2}},
~~~ Q = \pmatrix{ 1 & 0 & 0 \cr 0 & l_{2}/l_{1} & 0 \cr 0 & 0 & 1 }    .
\ee 
Note that this is indeed consistent with Eq. (\ref{esch}). 
The Maxwell equation in Eq. (\ref{emaxe}) possesses a symmetry under 
the change; $ g \rightarrow g^{'}= gh$ where $h$ is an arbitrary constant 
diagonal matrix. That is, $g^{'}$ again satisfies Eqs. (\ref{esch}) and 
(\ref{emaxe}). This symmetry becomes enhanced to a larger one when the 
oscillator strengths are equal ( $l_{1} = l_{2}= s $) so that $Q $ is the 
identity matrix. For the $\L$-system, the oscillator strength $s $ is positive 
which we assume throughout the paper. With equal oscillator strengths, 
$\tilde h$ can be a constant unitary matrix of the form
\be
{\tilde h} = 
\pmatrix{ h_{11} & h_{12} & 0 \cr h_{21} & h_{22} & 0 \cr 0 & 0 & h_{33} }.
\ee
In other words, $\tilde h$ is a constant matrix belonging to the group 
$ SU(2) \times U(1)$. In terms of physical variables, this symmetry amounts 
to the transform,
\be
\pmatrix{ \O^{'}_{1} \cr  \O^{'}_{2} } 
= \pmatrix{ h_{11}^{*} & h_{21}^{*} \cr h_{12}^{*} & h_{22}^{*} } 
\pmatrix{ \O_{1} \cr \O_{2}}, 
~~~ \pmatrix{ c^{'}_{1} \cr c^{'}_{2} \cr  c^{'}_{3} }
= \pmatrix{ h_{11}^{*} & h_{21}^{*} & 0  \cr h_{12}^{*} & h_{22}^{*} & 0 \cr 
0 & 0 & h_{33}^{*}} \pmatrix{ c_{1} \cr c_{2} \cr c_{3} } ,            
\label{sym}
\ee                  
where the primed variables are solutions of Eqs. (\ref{esch}) and 
(\ref{emaxe}) provided that unprimed variables are. In particular, 
if the unprimed solution is a single $2\pi $-pulse 
( $\O_{1} = 0, ~ \O_{2} \sim \mbox{sech}( t - x/v)/t_{p})$, the primed 
solution represents a simulton solution \cite{kono}. 
When the oscillator strengths are equal ($Q=1$), the 
theory becomes integrable and exact analytic solutions can be found. 
However, even for $Q \ne 1$, the Maxwell equation admits a Lax pair 
representation and infinite sequences of conserved integrals can 
be found \cite{park3}. In this paper, we assume the equal 
oscillator strengths so that the Maxwell equation becomes
\be
\pb (g^{-1} \pp g) = [T, ~  g^{-1} \Tb g] .
\label{emax2}
\ee
\section{B\"{a}cklund transformation}
The Maxwell equation in Eq. (\ref{emax2}) is equivalent to the 
consistency condition: $ [ L_{1}, ~ L_{2}] = 0$ of the overdetermined 
linear equations,
\ben
L_{1} \Psi &=& ( \pp +  g^{-1} \pp g + \l T )\Psi = 0 \nonumber \\
L_{2} \Psi &=& ( \pb + \li g^{-1} \Tb g )\Psi = 0  ,  
\label{lineq}
\een
where $\l $ is a spectral parameter. 
We may apply the inverse scattering method to Eq. (\ref{lineq}) and 
obtain exact solutions as in \cite{park2}. Instead, we present 
in this paper an alternative, simpler method - the B\"{a}cklund 
transformation (BT) - which allows a more direct construction of 
exact solutions. Moreover, the advantage of using the matrix potential 
$g$ becomes clear when the BT is used to establish a 
nonlinear superposition rule for a couple of single pulses. This is
similar to the linear case where the application of the usual 
superposition rule is easier in terms of the scalar potential rather 
than the electric field itself. Let $g_{0}$ and $\Psi_{0}$ be a 
particular solution of Eqs. (\ref{emax2}) and (\ref{lineq}), then 
$g_{1}$ is also a solution of Eq. (\ref{emax2}) if it satisfies 
the B\"{a}cklund transformation of type I (type I-BT):
\ben
\rm{\underline{Type ~ I-BT}}; ~~~~~~~~
 g^{-1}_{1} \pp g_{1} -  g^{-1}_{0} \pp g_{0} -i \eta 
[ g_{1}^{-1}g_{0}, ~ T] &=& 0 \nonumber \\
i \eta \pb (g_{1}^{-1}g_{0}) + g_{1}^{-1}\Tb g_{1} - g_{0}^{-1}\Tb g_{0}
&=& 0  .~~~~~~~~~~~~~~~~~~~~~~~~~~~~~~~~~~~~~~~~~~~~~~
\label{BT1}
\een   
Here, $\eta $ is an arbitrary parameter of the transformation. 
Once again, the type I-BT is a set of overdetermined first order partial 
differential equations whose consistency requires that $g_{0}$ and 
$g_{1}$ should be both solutions of Eq. (\ref{emax2}).
An equivalent expression of the BT is in terms of the linear function 
$\Psi$ as in \cite{shinnpb}. We define the B\"{a}cklund transformation of 
type II (type II-BT) by: 
\be
\rm{\underline{Type~ II-BT}} ;~~~~~~~~~~
\Psi_{1}={\l \over \l - i \eta }(1 + 
{i \eta \over \l }g_{1}^{-1}g_{0})\Psi_{0} .  
~~~~~~~~~~~~~~~~~~~~~~~~~~~~~~~~~~~~~~~~~~~~~~~~~~~~~
\label{BT2}
\ee       
It can be readily checked that $g_{1}$ and $\Psi_{1}$
satisfy Eqs. (\ref{emax2}) and (\ref{lineq}) provided Eq. (\ref{BT1}) 
holds and vice versa. The type II-BT is particularly useful for 
establishing a nonlinear superposition rule. Assume that 
$(g_{a}, ~ \Psi_{a}) $ and $(g_{b}, ~\Psi_{b})$ are two sets of solutions 
with BT parameters $\eta_{a}$ and $\eta_{b}$ respectively, 
solving the BT for a particular solution, $(g_{0}, ~ \Psi_{0})$. 
If we apply the BT once more to $(g_{a}, ~ \Psi_{a})$ with $\eta = \eta_{b}$ 
and also to $(g_{b}, ~ \Psi_{b})$ with $\eta = \eta_{a}$, and require 
that they result in the same solution (this amounts to the commutability 
of the diagram in Fig. 1), then we obtain from the type II-BT,
\ben
\Psi&=& {\l \over (\l - i \eta_{b}) }{\l \over (\l - i \eta_{a}) }(1 + 
{i \eta_{b} \over \l }g^{-1}g_{a})(1 + 
{i \eta_{a} \over \l }g_{a}^{-1}g_{0})\Psi_{0} \nonumber \\  
&=& {\l \over (\l - i \eta_{a}) }{\l \over (\l - i \eta_{b}) }(1 + 
{i \eta_{a} \over \l }g^{-1}g_{b})(1 + 
{i \eta_{b} \over \l }g_{b}^{-1}g_{0})\Psi_{0} .
\een
\unitlength 0.3cm
\begin{figure}
\begin{picture}(30,20)(-10,1)
\put(0,6){\framebox(4,2){$g_0, \J_0$}}
\put(12,12){\framebox(4,2){$g_a, \J_a$}}
\put(12,0){\framebox(4,2){$g_b, \J_b$}}
\put(24,6){\framebox(4,2){$g, \J$}}
\put(4,7){\vector(4,3){7.8}}
\put(16,13){\vector(4,-3){7.8}}
\put(4,7){\vector(4,-3){7.8}}
\put(16,1){\vector(4,3){7.8}}
\put(7,11){$\eta_a$}
\put(7,3){$\eta_b$}
\put(21,11){$\eta_b$}
\put(21,3){$\eta_a$}
\end{picture}
\caption{Commutability diagram for the nonlinear superposition rule.}
\end{figure}
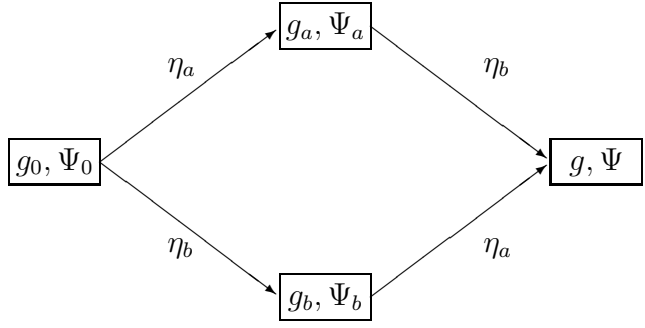
Equivalently, we have the nonlinear superposition of two solutions,
\be
g= (\eta_{b} g_{a} - \eta_{a} g_{b})g_{0}^{-1}(\eta_{b} g_{b}^{-1} 
-\eta_{a} g_{a}^{-1})^{-1} .     
\label{sup1}
\ee
Combining this expression with the type I-BT, we obtain a useful 
formula for the nonlinear superposition of the Rabi frequencies,
\be
g^{-1}\pp g = {1 \over 2}(g_{a}^{-1}\pp g_{a} + g_{b}^{-1}\pp g_{b} )
+  {i \over 2}[g^{-1}(\eta_{a} g_{b} +\eta_{b} g_{a}), ~ T] ,
\label{sup2}
\ee         
Or,
\ben
\O_{1} &=& {1 \over 2}(\O_{1}^{a} + \O_{1}^{b} ) - {i \over 2}F_{13} 
\nonumber \\
\O_{2} &=& {1 \over 2}(\O_{2}^{a} + \O_{2}^{b} ) - {i \over 2}F_{23}   
\nonumber \\
F &\equiv & (\eta_{b} g_{b}^{-1} -\eta_{a} g_{a}^{-1}) g_{0}   
(\eta_{b} g_{a} - \eta_{a} g_{b})^{-1}(\eta_{a} g_{b} +\eta_{b} g_{a}).
\label{sup3}
\een
\section{Matched pulses }
Now, we construct solutions by integrating the type I-BT directly. 
We choose the particular solution $g_{0}$ for a constant electric field.
\be
g^{-1}_{0} \pp g_{0} =   \pmatrix{ 0 & 0 & -i \O_{1} \cr      
0 & 0 & -i \O_{2} \cr - i\O_{1}^{*} & -i\O_{2}^{*} & 0} =
 \pmatrix{ 0 & 0 & -i\O_{0} \cr
                  0 & 0 & 0 \cr
                  - i\O_{0} & 0 & 0 }    \equiv \L 
\ee
for a real constant $\O_0$, so that 
\be
g_{0} =e^{\L z} =  \pmatrix{ \cos (\O_{0} z) & 0 & - i \sin (\O_{0} z) \cr
                  0 & 1 & 0 \cr
                  -i \sin (\O_{0} z ) & 0 & \cos (\O_{0} z) } .
\ee
If we set $f \equiv e^{-\L z}g_1$, the type I-BT becomes
\ben
f^{-1}\pp f + f^{-1}\L f -\L - i \eta [f^{-1} , ~ T] &=& 0  
\label{BT3-1}
\\
i \eta \pb f^{-1}  + f^{-1} \Tb f - \Tb &=& 0 .
\label{BT3-2}
\een
Since $f^{-1}\pp f$ is anti-hermitian, Eq. (\ref{BT3-1}) 
requires that 
\be
[f^{-1} - f, ~T] =0,
\ee
which we solve by taking 
\be
f^{-1} -  f = 2i \sin{\q}                               
\label{pro1}
\ee
for an arbitrary real parameter $\sin{\q }$. If we  rewrite $f$ 
in terms of another matrix $P$, 
\be
f \equiv e^{-\L z}g_1 = \cos{\q } (2 P -1) - i \sin{\q } ,
\label{pro2}
\ee
then Eqs. (\ref{pro1}) and (\ref{pro2}) imply that $P$ 
is a hermitian projection operator, i.e.
\be
P^{2} = P, ~~ P^{\dagger } = P .
\ee
In terms of $P$, Eqs. (\ref{BT3-1}) and (\ref{BT3-2}) can be 
written by 
\ben
(1 -P)( \pp + \L  -i \et T )P &=& 0 \nonumber \\
(1 -P)(i \et \pb - \Tb )P &=& 0     ,
\label{proj}
\een
where $ \et \equiv e^{i \q}\eta $. Since $P$ is a projection operator 
acting on the three-dimensional space, we may express $P$ using  
a three-dimensional vector $\vec{s}=(s_{1}, s_{2}, s_{3})$ by 
\be
P_{ij} =  s_{i} s_{j}^{*} / (\sum_{k=1}^{3} s_{k} s_{k}^{*})
\ee
which transforms Eq. (\ref{proj}) into a linear one
\ben
(\pp +\L-  i \et T) \vec{s} &=& 0 \nonumber \\
(i \et \pb - \Tb )\vec{s} &=& 0 .
\label{proj2}
\een
Since $\L $ and $\Tb $ commute, this may be integrated 
immediately resulting that
\be
s_{i} = \sum_{k=1}^{3}[\exp{\D}]_{ik}u_{k}, ~~~ \mbox{for}~~
\D = (i \et T - \L)z - {i \over \et }\Tb \zb
\ee
where $\vec{u}=(u_{1}, u_{2}, u_{3})$ is an arbitrary complex 
constant vector. Explicitly, 
\ben
s_{1} &=& (\cosh{\sqrt{K}z} + {\et  \over 2\sqrt{K}} 
\sinh{\sqrt{K}z} )u_{1} + {i \O_{0} \over \sqrt{K}} 
(\sinh{\sqrt{K}z}) u_{3} 
\nonumber \\
s_{2} &=& \exp({\et z \over 2} - { s\zb \over \et }) u_{2}
\nonumber \\
s_{3} &=& {i \O_{0} \over \sqrt{K}}( \sinh{\sqrt{K}z}) u_{1}  
+ (\cosh{\sqrt{K}z} - {\et  \over 2\sqrt{K}}\sinh{\sqrt{K}z})
u_{3}    , 
\label{svec}
\een
where $K = \et^{2} / 4 -\O_{0}^2 $.  
Finally, Rabi frequencies  are given by
\ben
\O_{1} &=& \O_{0} - 2 i \eta \cos{\q }s_{1}s_{3}^{*}/ 
(\sum_{k=1}^{3} s_{k} s_{k}^{*}) \nonumber \\
\O_{2} &=& -2 i \eta \cos{\q} s_{2}s_{3}^{*}/ 
(\sum_{k=1}^{3} s_{k} s_{k}^{*})  
\een
and probability amplitudes are obtained through Eqs. (\ref{gmat}) and 
(\ref{pro2}),
\ben
c_{1} &=& 2 \cos{\q} s_{1}s_{2}^{*}/ 
(\sum_{k=1}^{3} s_{k} s_{k}^{*})        \nonumber \\
c_{2} &=& \cos{\q} [2 s_{2}s_{2}^{*}/ 
(\sum_{k=1}^{3} s_{k} s_{k}^{*}) -1] + i \sin{\q} \nonumber \\
c_{3} &=& 2 \cos{\q} s_{3}s_{2}^{*}/ 
(\sum_{k=1}^{3} s_{k} s_{k}^{*})     .
\een
Note that this solution, combined with the symmetry transformation in 
Eq. (\ref{sym}), represents a rich family of single pulse solutions. 
For the vanishing $(\O_{0} = 0$), it becomes
\ben
\O_{1} &=& {1 \over N}(  - 2i \eta \cos{\q}u_{1}u_{3}^{*} \exp[i\eta z \sin{\q}] )
\nonumber \\
\O_{2} &=& {1 \over N}( - 2i \eta \cos{\q}u_{2}u_{3}^{*} \exp[i  (\eta z + {s \over \eta}\zb ) \sin{\q}
-{s\over \eta } \zb \cos{\q}] )   
\label{sol1}
\een
and
\ben
c_{1} &=& {1 \over N} ( 2 \cos{\q}u_{1}u_{2}^{*}\exp[(\eta z - {s \over \eta}\zb) \cos{\q}- 
i{s \over \eta}\zb \sin{\q}] )
\nonumber \\
c_{2} &=& {1\over N}( -|u_{1}|^2\exp[\eta z \cos{\q}-i \q]  +|u_{2}|^2  \exp[(\eta z - 
2{s \over \eta}\zb )\cos{\q} + i \q ] -|u_{3}|^2 \exp[-\eta z \cos{\q} - i \q ]  )
\nonumber \\
c_{3} &=& {1\over N}(2 \cos{\q}u_{3} u_{2}^{*} \exp[- {s \over \eta }\zb \cos{\q} 
-i (\eta z + {s \over \eta} \zb )\sin{\q} ]  )
\een
where
\be
N \equiv |u_{1}|^2 \exp[\eta z \cos{\q} ] +|u_{2}|^2 \exp[ \eta z \cos{\q} - 
{2 s\over \eta } \zb \cos{\q }] +|u_{3}|^2 \exp[- \eta z \cos{\q}] .
\ee
This solution for $u_1 \ne 0, u_2 \ne 0, ~ \eta \cos{\q} > 0$ describes the 
transfer of the $2\pi $-pump pulse in the limit $x \rightarrow - \infty$:
\ben
\O_{1} &\rightarrow& 0 , ~~ 
\O_{2} \rightarrow   - i \eta \cos{\q} \mbox{sech}\S_{1} 
e^{i\S_{2}} \nonumber \\
c_{1}  &\rightarrow& 0 , ~~ 
c_{2}  \rightarrow  {e^{\S_{1} + i \q } - e^{-\S_{1} -i \q } \over e^{\S_{1} } + e^{-\S_{1} }}
, ~~ 
c_{3}  \rightarrow  \cos{\q} e^{-i\S_{2}} \mbox{sech}\S_{1} 
\nonumber \\
\S_{1} &=& \cos{\q} ( \eta t - {(\eta^2 + s) \over c\eta }x + \phi_{1})   
, ~~~
\S_{2} =   \sin{\q} ( \eta t - {(\eta^2 - s) \over c\eta }x + \phi_{2}) 
\label{pump}
\een
to the $2\pi $-Stokes pulse moving with the velocity of light in the limit 
$x\rightarrow \infty$;
\ben
\O_{1} &\rightarrow&  - i \eta \cos{\q} 
\mbox{sech}[ \cos{\q}  \eta (t - {x \over c}) ]
\exp[i  \sin{\q}  \eta (t - {x \over c}) ]  , ~~ 
\O_{2} \rightarrow  0 \nonumber \\
c_{1} &\rightarrow & 0 , ~~ 
c_{2} \rightarrow  -e^{-i \q} , ~~ 
c_{3} \rightarrow  0 ,
\label{stoke}
\een 
where the arbitrary constants $\phi_{i}, i=1,2$ determine soliton 
positions in time and space.
In the case of $\q =0$, this transfer of 
$2\pi $-pulse has been given in \cite{bolshov2}. For $u_1=0$ or $u_2=0$, 
the solution remains as the steady state $2 \pi $-pulse given in 
Eqs. (\ref{pump}) or (\ref{stoke}) respectively without causing any 
transfer of pulses. This steady pulse, in connection with the symmetry 
transform in Eq. (\ref{sym}), is the simulton solution \cite{kono}. 
The free parameter $\q $ measures the amount of self-detuning of a 
pulse from the carrier frequency. The $2\pi $-Stokes pulse in 
Eq. (\ref{stoke}) is the same as the usual $2\pi $-pulse ($\q =0$) but 
with the carrier frequency shifted by the amount $\d w =\eta \sin{\q}  $. 
On the other hand, Eq. (\ref{pump}) shows that the  $2\pi $-pump pulse 
receives a time independent phase factor $\exp[ i (s\sin{\q}/c\eta) x]$ 
in addition to the shift of carrier frequency. We emphasize that this 
detuning has nothing to do with the frequency detuning of electromagnetic 
fields from the resonance line. In fact, our system is on resonance and 
thus the parameter $\q$ measures the self-generated detuning 
of each pulses. The effect of $\q $ to a single $2\pi $-pulse is to broaden 
the pulse shape maintaining the $2\pi $ area of the envelope which 
is adjusted by the shift of the carrier frequency. 
Recall that due to the symmetry in Eq. (\ref{sym}), a more general 
expression for a single pulse arises as a linear mixture of $\O_{1}$ and 
$\O_{2}$ in Eq. (\ref{sol1}) which possess a wide range of free parameters.

If $\O_{0} \ne 0$, the solution 
describes pulses more general than the hyperbolic-secant type. 
For the simplicity of analysis, we assume that $\q = 0$ and 
$|\eta|\ge 2 |\O_{0}|$. We also restrict to the parameters; $u_{1}=r_{1}, ~
u_{2}=r_{2} , ~ u_{3}= i r_{3}$ for real $r_{i}$ and rewrite 
Eq. (\ref{svec}) for the notational convenience as follows;
\ben
s_1 &=& A \exp(\sqrt{K} z) + B \exp(-\sqrt{K} z) \nonumber \\ 
s_2 &=& \sqrt{A^2 + C^2} \exp({\eta z \over 2}-{s \zb \over \eta}) 
\nonumber \\
s_3 &=& i C \exp(\sqrt{K} z) + iD  \exp(-\sqrt{K} z) 
\een
where $r_{2}$ is chosen to be $\sqrt{A^2 + C^2}$ by an appropriate 
choice of the coordinate origin. The coefficients are defined by
\ben
A&=& {1 \over 2}( 1 + {\eta \over 2 \sqrt{K}})r_{1} - 
{\O_{0} \over 2 \sqrt{K}}r_{3} \nonumber \\
B&=& {1 \over 2}( 1 - {\eta \over 2 \sqrt{K}})r_{1} +
{\O_{0} \over 2 \sqrt{K}}r_{3} \nonumber \\
C&=& {\O_{0} \over 2 \sqrt{K}}r_{1} + 
{1 \over 2}( 1 - {\eta \over 2 \sqrt{K}})r_{3} 
 \nonumber \\   
D&=& -{\O_{0} \over 2 \sqrt{K}}r_{1} + 
{1 \over 2}( 1 + {\eta \over 2 \sqrt{K}})r_{3} 
\nonumber \\
K &=& {1 \over 4} \eta^2 - \O_{0}^2 . 
\label{before}
\een
In the limit where $x \rightarrow -\infty $,
the solution takes an asymptotic form:
\be
\O_{1} = -\O_{0} \mbox{tanh} \S_{I}, ~~ 
\O_{2} = {\eta C \over \sqrt{A^2 + C^2 }}\mbox{sech} \S_{I} 
\label{counter1}
\ee
and
\ben
c_{1} &=& -{\eta C \over \sqrt{A^2 + C^2 }} \mbox{sech} \S_{I} , ~ ~ 
c_{2} =- \mbox{tanh} \S_{I} , ~~ 
c_{3} = { i C \over \sqrt{A^2 + C^2 }} \mbox{sech} \S_{I} 
\nonumber \\
\S_{I} &=& {2s + (\eta^2 - 2\eta \sqrt{K}) \over 
2\eta c} (x - v_{I}t) 
\label{counter2}  
\een
where the velocity $v_{I}$ is
\be
v_{I} = { \eta^2 - 2\eta \sqrt{K} \over  
2s + \eta^2 - 2 \eta \sqrt{K} }c ,
\ee
which is less than the light velocity  $c$.  Comparison of electric fields in 
Eq. (\ref{counter1}) with initial populations in Eq. (\ref{counter2}) shows 
that two laser pulses are arranged in the so-called counterintuitive 
order \cite{count}.
In the $x \rightarrow \infty $ limit, the asymptotic form of the 
solution is 
\be
\O_{1} = -\O_{0} \mbox{tanh} \S_{F}  + \O_{1}^{S} , ~~ 
\O_{2} = {\eta C \over \sqrt{A^2 + C^2 }}
\mbox{sech} \S_{F}
\label{after1}    
\ee
and
\be
c_{1} = {B \over \sqrt{B^2 + D^2 }}\mbox{sech} \S_{F} , ~~ 
c_{2} = -\mbox{tanh} \S_{F} , ~~ 
c_{3} ={iD \over \sqrt{B^2 + D^2 }}\mbox{sech} \S_{F} 
\ee                                          
where
\ben
\S_{F} &=&  {2s + (\eta^2 + 2 \eta \sqrt{K}) \over 2\eta c} (x - v_{F}t) -\D_{0} 
\nonumber \\   
v_{F} &=& { \eta^2 + 2 \eta \sqrt{K} \over 
2s + \eta^2 + 2 \eta \sqrt{K} }c 
\nonumber \\        
\D_{0} &=& {1\over 2} \ln {B^2 + D^2 \over A^2 + C^2 }
\nonumber \\
\O_{1}^{S} &\equiv &  {  \O_{0} (r_{1}^{2} + r_{3}^{2}) 
- \eta r_{1}r_{3} \over (A^2 + C^2 )e^{\D_{0}}\mbox{cosh} 
(2 \sqrt{K} (t-x/c)-\D_{0}) + AB + CD} .
\een
In the far past, this solution represents a matched-solitary-wave 
pair (MSP) moving with velocity $v_{I}$ whose invariant shape is not 
the hyperbolic-secant type. Eq. (\ref{after1}) shows that this MSP 
in general breaks up into another MSP with slower velocity 
$(v_{F} < v_{I})$ and a soliton pulse moving with light velocity. 
Figure 2 shows explicitly this breakup behavour with parameters, 
$s=1$, $r_1 =0.7$, $r_3 =1$, $\O_0 = 0.5$ and $\eta=1.5$.


By changing the sign of $\eta $, one could equally consider the 
reverse process of fusing a MSP and a soliton into another MSP.
In the specific case where $K=\eta^2 /4 -\O_{0}^2 $ vanishes, 
this solution reduces to the steady state pulse which agrees with
the MSP solution by Hioe and Grobe \cite{hioe},
\be
\O_{1} = -\O_{0} \mbox{tanh} \S_{S} , ~~ 
\O_{2} = \sqrt{2} \O_{0} \mbox{sech} \S_{S}  
\ee
and
\be
c_{1} = {1\over \sqrt{2}} \mbox{sech} \S_{S} , ~~ 
c_{2} = - \mbox{tanh} \S_{S} , ~~ 
c_{3} =  {i \over \sqrt{2}} \mbox{sech} \S_{S}
\ee
where 
\ben
\S_{S} &=& {s + 2\O_{0}^2 \over 2\O_{0} c}(x-v_{S}t) , ~~
v_{S} = 2\O_{0}^2 /(s + 2 \O_{0}^2 ) .
\label{velo}
\een
Thus, our solution generalizes the result of Hioe and Grobe and 
shows that MSP propagates steadily only when the velocity $v_{S}$ of MSP 
is specifically given by the parameter $\O_{0}$ as in Eq. (\ref{velo}). 
Otherwise, it breaks up as explained before.
In other words, in the case of MSP where $K\ne 0$, the hyperbolic-tangent 
pulse $\O_{1}$ induces a partial transfer of the other soliton pulse 
$\O_{2}$ into the light velocity one. This is similar to the behavior of 
adiabaton moving with a slower velocity which also generates a light velocity 
solitary pulse during its formation \cite{adiabaton}.\footnote{We thank 
Referee for raising this point.}
where the most general expression of 
our MSP solution arises when we incorporate the $SU(2) \times 
U(1)$ symmetry transform as in Eq. (\ref{sym}) as well as the 
self-detuning effect $(\q \ne 0)$. This solution contains a large 
number of free parameters which account for the variety of pulse shapes 
and initial conditions required for the atoms. The stability of this 
MSP solution against small perturbations is an important issue \cite{Infeld} 
which we plan to consider elsewhere.
\section{Invertons and breathers} 
The nonlinear superposition rule given in Eqs. (\ref{sup1})-(\ref{sup3}) 
allows the superposition of two MSP solutions $g_{a}$ 
and $g_{b}$ whenever they are obtained from $g_{0}$ with BT 
parameters $\eta_{a}$ and $\eta_{b}$ respectively. In general, $g_{a}$ 
and $g_{b}$ could possess different set of free parameters and the 
superposition rule requires only that they are obtained from the same 
$g_{0}$ through BT. If $g_{a}$ and $g_{b}$ have same set of parameters 
except for BT parameters, we could analytically continue the BT 
parameter in Eq. (\ref{sup1}) to the complex case in such a way that
\be
\eta_{a} = \exp[i \q_{B}] , ~~ \eta_{b} = \exp[- i \q_{B}]
\ee
for a certain real parameter $\q_{B}$, and that $g$ is still unitary 
and becomes a nonabelian breather solution. As an example,  we take 
the one-soliton to be the MSP in the previous section with a set of 
parameters
\be
\O_{0} = 0, ~ u_{1} =1, ~ u_{2} = 1  , u_{3} = - i .
\ee
Then, the nonabelian breather solution has an asymptotic form for 
$x \rightarrow - \infty$, 
\ben
\O_{1} &=& 0 
\nonumber \\
\O_{2} &=& { \sin{2 \q_{B}} (\exp[(2x/c -t) \cos{\q_{B}}] \sin(\q_{B} 
+ t \sin{\q_{B}}) + \exp[(t - 2x/c)\cos{\q_{B}}] \sin(\q_{B} - 
t\sin{\q_{B}}) ) \over
-1 + \cos^{2}{\q_{B}}\cos(2t \sin{\q_{B}}) - \cosh[(4x/c - 2t)
\cos{\q_{B}} ]\sin^{2}{\q_{B}}}    
\nonumber \\
\een
and for $x \rightarrow \infty $,
\ben 
\O_{1} &=&  { \sin{2\q_{B}} ( e^{Z_{C}} \sin(\q_{B} - Z_{S}) 
+ e^{-Z_{C}} \sin (\q_{B} + Z_{S})) \over
-1+ \cos^{2}{\q_{B}} \cos(2 Z_{S}) -\cosh(2 Z_{C}) \sin^{2}{\q_{B}}}  
\nonumber \\
\O_{2} &=& 0   
\nonumber \\                    
Z_{C} &=& (t -x/c) \cos{\q_{B}}, ~~ Z_{S} = (t -x/c) \sin{\q_{B}}  .
\een
This describes a breathing $0 \pi$-pulse $\O_{2} ~(v < c )$ 
which transfers to the nonbreathing $\O_{1}$ pulse moving with the 
velocity of light. This is a typical motion of a 
nonabelian breather. Figure 3 shows a breathing motion
with $\theta_B =1.2$. Different values of parameters $u_{i}$ in 
general distort the shape of $0 \pi$-pulse so that the time area 
of the pulse is nonzero, but they lead to the essentially same 
type of transferring motion except for the case $u_{1}=0$ or 
$u_{2}=0$ where it becomes a steady state breather.


For real BT parameters $\eta_{a}$ and $\eta_{b}$, the superposed 
solution $g$ in general describes the scattering of two MSP solutions. 
Here, we concentrate only on a particular case of two soliton 
solutions, which we name as a pair of ``invertons" due to their 
interesting new features. Consider one solitons $g_{a}$ and $g_{b}$ 
given by
\ben
g_{a} &=& \pmatrix{-1 & 0 & 0 \cr
                  0 & \cos{\phi_{a}} & \sin{ \phi_{a}} \cr
                  0 & \sin{ \phi_{a}} & - \cos{\phi_{a}} } 
\nonumber \\ 
\cos{\phi_{a}} &=& \mbox{tanh} [\eta_{a} t - {\eta_{a}^{2} + s 
\over c \eta_{a} }x] 
\nonumber \\
\sin{ \phi_{a}} &=& \mbox{sech} [\eta_{a} t - {\eta_{a}^{2} + s 
\over c \eta_{a} }x] 
\een
and 
\ben
g_{b} &=& \pmatrix{\cos{\phi_{b}}  & 0 &  \sin{ \phi_{b}} \cr
                  0 & -1 & 0 \cr
                  \sin{ \phi_{b}}  & 0 & - \cos{\phi_{b}} } 
\nonumber \\ 
\cos{\phi_{b}} &=& \mbox{tanh} [\eta_{b} (t - x/c) ] 
\nonumber \\
\sin{ \phi_{b}} &=& \mbox{sech} [ \eta_{b} (t - x/c) ] .
\een
In terms of Rabi frequencies, these correspond to 
\be
\O^{(a)}_{1} = 0 , ~~ \O^{(a)}_{2} = 
- i \eta_{a} \mbox{sech}[ \eta_{a} t - {\eta_{a}^{2} + s 
\over c \eta_{a} }x]   
\ee
and
\be
\O^{(b)}_{1} = - i  \eta_{b} \mbox{sech}[ \eta_{b} (t - x/c) ]   
, ~~   \O^{(b)}_{2} = 0      ,
\ee
that is, they represent two $2\pi $-pulses with 
different resonance frequencies, one moving with light velocity and 
the other moving with slower velocity.


The nonlinear superposition rule gives rise to the superposed 
Rabi frequencies and probabilities such that
\ben
\O_{1} &=& - i \eta_{b} \sin{\phi_{b}}{\eta^{2}_{a}\cos{\phi_{a}} 
+ \eta_{a} \eta_{b} (1+ \cos{\phi_{a}}) + \eta^{2}_{b} \over
\eta_{a}^{2} + \eta_{b}^{2} + \eta_{a}\eta_{b} (1 + \cos{\phi_{a}} 
+\cos{\phi_{b}} - \cos{\phi_{a}}\cos{\phi_{b}})}
\nonumber \\
\O_{2} &=& -  i \eta_{a} \sin{\phi_{a}}{\eta^{2}_{a} + 
\eta_{a}\eta_{b} (1+ \cos{\phi_{b}} ) + \eta^{2}_{b} \cos{\phi_{b}} \over
\eta_{a}^{2} + \eta_{b}^{2} + \eta_{a}\eta_{b} (1 + \cos{\phi_{a}} 
+\cos{\phi_{b}} - \cos{\phi_{a}}\cos{\phi_{b}})}  
\een
and
\ben
c_{1} &=& { (\eta_{a} + \eta_{b})\eta_{b} \sin{\phi_{a}} \sin{\phi_{b}} 
\over
\eta_{a}^{2} + \eta_{b}^{2} + \eta_{a}\eta_{b} (1 + \cos{\phi_{a}} 
+\cos{\phi_{b}} - \cos{\phi_{a}}\cos{\phi_{b}})}  
\nonumber \\
c_{2} &=& {- (\eta_{a}^{2} + \eta_{b}^{2}) \cos{\phi_{a}} 
-\eta_{a}\eta_{b} (1 + \cos{\phi_{a}} 
-\cos{\phi_{b}} + \cos{\phi_{a}}\cos{\phi_{b}}) 
\over
\eta_{a}^{2} + \eta_{b}^{2} + \eta_{a}\eta_{b} (1 + \cos{\phi_{a}} 
+\cos{\phi_{b}} - \cos{\phi_{a}}\cos{\phi_{b}})}  
\nonumber \\
c_{3} &=& {- (\eta_{a}^{2} + \eta_{a}\eta_{b}(1+ \cos{\phi_{b}}) 
+ \eta_{b}^{2}  \cos{\phi_{b}}) \sin{\phi_{a} } 
\over
\eta_{a}^{2} + \eta_{b}^{2} + \eta_{a}\eta_{b} (1 + \cos{\phi_{a}} 
+\cos{\phi_{b}} - \cos{\phi_{a}}\cos{\phi_{b}})}  .
\een
This solution describes the scattering of two invertons where the 
fast moving inverton $I_{F}$ passes through the slow moving one 
$I_{S}$. Before the collision, their asymptotic forms are given 
by the following configurations;
\ben
\underline{\mbox{Inverton}~ I_{F}}  &;&  \cos{\phi_{a}} = 1
\nonumber \\
&& c_{1} = 0, ~ c_{2} = -1, ~ c_{3}  =0 
\nonumber \\
&& \O_{1} = - i \eta_{b} \mbox{sech}[\eta_{b}(t -x/c)] , ~~ \O_{2} = 0 
~~~~~~~~~~~~~~~~~~~~~~~~~~~~~~~~~~~~~~~~~~~~~~~~~~~~~~
\een
and
\ben  
\underline{\mbox{Inverton}~ I_{S}} &;& \cos{\phi_{b}} = 1
\nonumber \\
 && c_{1} = 0, ~ c_{2} = -\tanh[\eta_{a} t - {\eta_{a}^{2} + s 
\over c \eta_{a} }x], ~ c_{3}  = -\mbox{sech}[\eta_{a} t - 
{\eta_{a}^{2} + s \over c \eta_{a} }x]  
\nonumber \\
&& \O_{1} = 0, ~~ \O_{2} = - i \eta_{a} \mbox{sech}[\eta_{a} t - 
{\eta_{a}^{2} + s \over c \eta_{a} }x] . 
~~~~~~~~~~~~~~~~~~~~~~~~~~~~~~~~~~~~~~~~~~~~~~~~~~
\een   
After the collision, their asymptotic forms are
\ben
\underline{\mbox{Inverton}~ I_{F}} &;&  \cos{\phi_{a}} = -1
\nonumber \\
&& c_{1} = 0, ~ c_{2} = 1, ~ c_{3}  =0 
\nonumber \\
&& \O_{1} = i  \eta_{b} \mbox{sign}({\eta_{a} + \eta_{b} \over 
\eta_{a} - \eta_{b}}) \mbox{sech}[\eta_{b}(t -x/c) + \d ] , 
~~ \O_{2} = 0  ~~~~~~~~~~~~~~~~~~~~~~~~~~~~~~~~~~~~~
\een   
and
\ben
\underline{\mbox{Inverton} ~I_{S}} &;& \cos{\phi_{b}} = -1
\nonumber \\
 && c_{1} = 0, ~ c_{2} = -\tanh[\eta_{a} t - {\eta_{a}^{2} + s 
\over c \eta_{a} }x + \d ] , \nonumber \\
&& c_{3}  = -\mbox{sign} 
({\eta_{a} + \eta_{b} \over \eta_{a} - \eta_{b}}) 
\mbox{sech}[\eta_{a} t - {\eta_{a}^{2} + s \over c \eta_{a} }x + \d ]  
\nonumber \\
&& \O_{1} = 0, ~~ \O_{2} = - i \eta_{a} \mbox{sign} 
({\eta_{a} + \eta_{b} \over \eta_{a} - \eta_{b}})
 \mbox{sech}[\eta_{a} t - {\eta_{a}^{2} + s \over c \eta_{a} }x + \d ] ,
 ~~~~~~~~~~~~~~~~~~~~~~~~~~
\een   
where sign denotes the sign function and the phase shift parameter $\d $ 
is
\be
\d =  \mbox{ln}|{\eta_{a} + \eta_{b} \over \eta_{a} - \eta_{b}}| .
\ee                 
The appearance of the sign function is a novel feature of soliton
scatterings in a three-level $\L $ system which does not arise in the 
two-level atom case. A careful analysis of the above asymptotic 
solutions shows that the broader pulse changes its sign after the 
collision (thereby the name inverton) while the sharper pulse remains 
the same. Also, the phase shift always arises in such a way that the 
slow inverton receives time retardation while the fast one recieves 
time advance. This implies that the fast inverton moving with light 
velocity moves faster than light through the scattering process!
However, this does not violate causality. In fact, the fast 
inverton has infinitely stretched tails which trigger the slow inverton 
to transfer its energy to the fast one.

Figure 4 shows the scattering 
of the invertons with parameters $\eta_a =0.5$, $\eta_b =0.9$ and $s=1$. 
In order to see that causality is not violated, we consider the energy 
conservation law given by 
\be
{\pp \over \pp t } \r = - {\pp \over \pp x} j
\ee 
where
\ben
\r &=& |\O_1 |^2 + |\O_2 |^2 +|c_1 |^2 + |c_2 |^2
  =|\O_1 |^2 + |\O_2 |^2 +1 -|c_3 |^2 
\nonumber \\
j &=& c (|\O_1 |^2 + |\O_2 |^2 ).
\een


Figure 5 shows the energy transfer process between the two invertons. 
When the right end tail of the fast inverton $I_{F}$ reaches the slow 
inverton $I_{S}$, the energy is transferred from
$I_{S}$ to $I_{F}$ and $I_{F}$  receives phase advance. After the center
of $I_{F}$ reaches $I_{S}$, the energy of $I_{F}$ is absorbed into the 
$I_{S}$ which results in a retarded phase shift of $I_{S}$. 
Thus, no energy is really transferred faster than light. 
The scattering process is indeed 
a nicely balanced exchange process where the scattered invertons 
maintain their shapes invariant except for the change of sign.  
The change of polarizations of simultons in the scattering process 
has been known earlier \cite{pola} and the scattering of two 
single-frequency $2\pi $-pulses with different resonance frequencies 
is also considered in \cite{bolshov2}. However, the binary, sign 
changing behavior of invertons is a new aspect which did not receive 
attention in earlier works, and hopefully will find some applications 
in the future.

\section{Discussion}  
In this paper, we have introduced a new formalism for the three-level 
coupled Sch\"{o}dinger-Maxwell equation based on a matrix potential variable, 
and through the B\"{a}cklund transformation we obtained various new solutions 
which generalize previously known solutions. We found new MSP solutions, 
more general than the steady state MSP found by Hioe and Grobe, which 
account for the generic breakup behavior of a MSP into another 
MSP and a light velocity soliton. Two-soliton solutions and nonabelian 
breathers are obtained through the nonlinear superposition rule and 
scattering of invertons are analyzed in detail. In this new formalism, 
we have demonstrated the $SU(2) \times U(1)$ group symmetry of the 
three-level system with equal oscillator strengths and using that we 
have found a general expression of solutions which mixes pulses with 
different velocities. This group symmetry also accounts for the variety 
of steady state MSP solution by Hioe and Grobe which linearly superpose 
different Jacobi elliptic functions. Another new feature of our formalism 
is the introduction of a self-detuning parameter $\q$ to the solutions. 
For each $2 \pi$-pulses, these $\q$-values are conserved in time and 
survives from the scattering process. That is, it is a conserved charge  
which can be used to label each pulses in addition to the area of pulses. 
We also emphasize that our new formalism is not restricted to the 
three-level case but also can be extended to other multi-level 
cases \cite{park2} with the same analytic power.  

Finally, we comment on matched pulse propagation through absorbing media.
In this case, the time rate equation for the probability amplitude $c_{3}$ 
in Eq. (\ref{Sch}) is replaced by $\pp c_{3} /\pp t \rightarrow  
\pp c_{3} /\pp t + \gamma c_{3} $, which implies the decaying of the 
excited state $|3>$ at a rate $\gamma$ to states other than $|1>$ and $|2>$.
This obviously breaks the probability conservation law; $|c_{1}|^2 + |c_{2}|^2
+ |c_{3}|^2 =1$ so that our matrix potential formalism does not apply 
to this case. Nevertheless, it is important to note that the 
$SU(2) \times U(1)$ group symmetry of the three-level system with equal 
oscillator strengths persists even in the case of an absorbing medium.
This shows that recent analytic and numerical studies on matched pulse 
propagation through absorbing media \cite{harr}-\cite{Vem2} can be extended 
to more general initial conditions. For instance, most studies assume the 
standard initial condition of the population being in the ground state. The 
$SU(2) \times U(1)$ group symmetry as indicated in Eq. (\ref{sym}) maps the 
standard initial condition to a coherently prepared one which linearly 
superposes $|1>$ and $|2>$. This also maps two Rabi frequencies 
$\O_{1}$ and $\O_{2}$ via the linear transform as in Eq. (\ref{sym}) so that 
one obtains a more general description of pulse propagation. Thus, for 
example, when a matched pulse of equal amplitude is found, the above 
symmetry generates a set of new solutions with unequal amplitudes. 
This agrees with results in the recent numerical study \cite{Vem2}, 
which showed the existence of stable, shape invariant matched pulses of unequal 
amplitudes depending on the coherent preparation of initial conditions.

\vglue .2in

 {\bf ACKNOWLEDGEMENT}
\vglue .2in
This work was supported in part by the program of Basic Science Research, 
Ministry of Education BSRI-97-2442, and by Korea Science and Engineering 
Foundation through CTP/SNU and  971-0201-004-2.

\vglue .2in

\end{document}